\documentclass{tmsce}

\usepackage[normalem]{ulem}
\usepackage{natbib}
\usepackage{multirow}
\usepackage{xcolor}
\usepackage{lipsum}
\usepackage{amsmath,amsfonts,amsthm}
\usepackage{algorithmic}
\usepackage{algorithm}
\usepackage{array}
\usepackage[caption=false,font=normalsize,labelfont=sf,textfont=sf]{subfig}
\usepackage{textcomp}
\usepackage{stfloats}
\usepackage{url}
\usepackage{verbatim}
\usepackage{graphicx}
\usepackage{cite}
\usepackage{mwe} 

\usepackage[dvipsnames,table]{xcolor}
\definecolor{LightCyan}{rgb}{0.88,0.95,1}

\doi{}

\title{Don't Disregard the Data for Lack of a Likelihood: Bayesian Synthetic Likelihood
for Enhanced Multilevel Network Meta-Regression}

\authors{Harlan Campbell$^{1,2}$,  Charles C. Margossian$^{1}$, Jeroen P. Jansen$^{2,3}$, Paul Gustafson$^{1}$}
\affiliation{$^{1}$Department of Statistics, University of British Columbia, Vancouver, Canada\\
$^{2}$HEOR - Precision AQ\\
$^3$Department of Clinical Pharmacy, University of California, San Francisco\\
\textit{Corresponding author:} harlan.campbell@stat.ubc.ca}

\received{}
\revised{}
\accepted{}


\begin{document}
\maketitle

\begin{abstract}
Multilevel network meta-regression (ML-NMR) enables population-adjusted indirect treatment
comparisons by combining individual patient data (IPD) with aggregate data. When individual-level
covariates are unavailable, ML-NMR marginalizes over the covariate distribution, but this strategy cannot
exploit subgroup-level summary results that are often available and potentially highly informative. We
propose using Bayesian Synthetic Likelihood (BSL) to leverage this ancillary summary information
{\color{black} and present an implementation strategy for Hamiltonian Monte Carlo (HMC), a gradient-based Markov chain Monte Carlo (MCMC) algorithm.}
At each MCMC iteration, the BSL method imputes missing covariates by sampling from the model-implied conditional distribution, computes synthetic subgroup summaries from the imputed data, and matches these synthetic summaries to observed summaries via a multivariate normal synthetic likelihood.
{\color{black}
Fitting this model with HMC presents multiple challenges:
first, gradients cannot be computed exactly but must be estimated stochastically; and second, the model's likelihood may be non-differentiable at certain points, a pathology that can deeply frustrate the performance of HMC.
We address these challenges with pre-drawn random numbers,  continuous relaxation of the likelihood, and Pareto-smoothed importance sampling.
}
This work (1) introduces a novel application of BSL to missing data problems where summary statistics from the complete dataset are available despite substantial missingness in the individual-level data, (2) demonstrates how BSL strategies can be implemented within Stan's HMC framework, and (3) shows, using a network of plaque psoriasis trials, that BSL-enhanced ML-NMR can substantially improve upon standard ML-NMR by leveraging informative ancillary information.
\end{abstract}

\printkeywords
\printdates

\tmsceendfrontmatter


\section{Introduction}

Health policy and reimbursement decisions increasingly require unbiased
estimates of relative treatment effectiveness for specific patient
populations. When head-to-head trials are unavailable, network
meta-analysis (NMA) synthesizes evidence across multiple studies
comparing different treatment pairs. However, standard aggregate-data NMA
can produce biased estimates when effect-modifying covariates are
distributed differently across study populations. This has motivated the development of population adjustment methods that explicitly account for covariate
distributions and treatment-by-covariate interactions \citep{phillippo2016nice, phillippo2018methods}.

\citet{phillippo2020multilevel} proposed multilevel network
meta-regression (ML-NMR), a Bayesian framework that combines individual
patient data (IPD) from some studies with aggregate data from others by
integrating an individual-level outcome model over each study's
covariate distribution, with posterior inference conducted via Markov
chain Monte Carlo (MCMC).
When individual-level covariates are unavailable, ML-NMR marginalizes
over the covariate distribution, avoiding aggregation bias and enabling principled use of mixed evidence. This strategy is based on the earlier proposal of \citet{jansen2012network}, who modeled aggregate-data studies as weighted averages of subgroup-specific outcomes to avoid ecological bias when combining individual patient data with aggregate data.

ML-NMR is currently regarded as the state-of-the-art for population-adjusted NMA with partial IPD \citep{phillippo2023validating}.  However, the marginalization strategy of ML-NMR cannot exploit all of the available evidence. Consider a typical scenario in health technology assessment: a published trial reports individual-level outcomes and treatment  assignments (e.g., the number of individuals in the treatment and control arms who do and do not experience disease progression) but withholds individual-level covariates (e.g., the age, sex and baseline disease severity of each individual) due to privacy or
proprietary concerns. The same publication does, however, include standard subgroup analyses: odds ratios for treatment effects among patients with baseline disease severity above versus below the median, or
stratified by binary risk factors such as sex or prior treatment history \citep{wang2007statistics, sun2011influence}. ML-NMR handles this scenario by integrating over the covariate distribution at the individual level, and since the marginalized likelihood has no natural place to condition on subgroup-level contrasts,
the subgroup summaries are effectively ignored. This is unfortunate as these summaries are far from uninformative: they encode direct evidence about how treatment effects vary with patient covariates, which is precisely the effect
modification that population adjustment seeks to estimate. They may also be informative less obvious ways. \citet{huang2025utilizing} recently demonstrated that routinely reported subgroup data can improve heterogeneity estimation when few trials are available.  More broadly,  discarding the subgroup summaries represents a potentially substantial loss of information \citep{sun2010subgroup, gabler2009dealing}.

We propose addressing this gap by extending ML-NMR to incorporate
subgroup summary information using Bayesian Synthetic Likelihood (BSL), a likelihood-free inference (LFI) method \citep{wood2010statistical, price2018bayesian} that has been applied successfully in a range of complex Bayesian modelling settings \citep{picchini2019bayesian, an2022bsl}. BSL approximates the intractable likelihood of
summary statistics by generating ``synthetic'' datasets, computing
synthetic summaries, and comparing them to observed summaries via a
multivariate normal likelihood. In our setting, when individual
covariate values are missing but subgroup summaries are available, BSL
generates synthetic complete datasets at each MCMC iteration by imputing
covariates from the model-implied conditional distribution, computes
synthetic subgroup statistics from these imputations, and uses the
resulting synthetic likelihood to guide inference toward parameter values
consistent with both the individual-level data, aggregate-level data, and the published subgroup results.

Our approach is related to, but distinct from, the method of Network
Meta-Interpolation (NMI) proposed by \citet{harari2023network}, which
also exploits subgroup analyses to adjust for effect modification in NMA.
NMI directly transforms treatment effect estimates at the subgroup and
study level into estimates at covariate values common to all studies.
However, this transformation is only approximate, and NMI has been
criticized for potential bias arising from mixing incompatible estimates
within the model \citep{phillippo2025effect}. In contrast, our approach embeds the subgroup information within a coherent generative model through the synthetic likelihood, preserving the principled individual-level framework of ML-NMR.

Despite the fact that subgroup analyses are routinely reported in
publications, clinical study reports, and regulatory submissions
\citep{wei2024description}, a principled approach for leveraging this
information within population-adjusted indirect treatment comparisons
has been lacking. This paper makes three main contributions.

First, we introduce a novel application of BSL to missing data problems where summary statistics computed from the complete dataset are available despite substantial missingness in the individual-level data, a scenario often described in the evidence synthesis literature \citep{riley2010meta} but not previously addressed in the BSL literature. We develop this idea in Section~\ref{sec:sec2} through a simple missing data problem that illustrates the core BSL mechanism.

\textcolor{black}{
Second, we demonstrate how BSL can be implemented in a probabilistic framework that uses Hamiltonian Monte Carlo (HMC) \citep{neal2012mcmc, betancourt2017hmc}.
HMC is a popular and battle-tested gradient-based MCMC algorithm, which scales well in high-dimensions.
As a result, HMC has emerged over the last decade as the main inference engine for modern probabilistic programming languages \citep{strumbelj2024software}, including Stan \citep{carpenter2017stan, stanmanual2025}, PyMC \citep{pymc2023}, and TensorFlow Probability \citep{dillon2017tensorflow_distributions}.
Stan itself is a natural choice for BSL as ML-NMR is already implemented in Stan via the \texttt{multinma} R package \citep{phillippo2023r}.
However, standard implementations of HMC \citep[e.g.,][]{hoffman2014nuts, betancourt2017hmc}, such as the one in Stan, impose several constraints on the Bayesian model that users can specify.
For BSL, the two relevant constraints are: (i) the likelihood must be a deterministic function of the data and model parameters; and (ii) the likelihood must be differentiable.
Unfortunately, both conditions are violated by BSL.
In Section~\ref{sec:sec3}, we address these challenges by substituting the exact (intractable) likelihood with an approximate likelihood, obtained with ``common random numbers'' \citep{meeds2015hamiltonian} and a continuous relaxation.
We then correct for this approximation in a post-sampling step, using Pareto-smoothed importance sampling (PSIS) \citep{vehtari2024pareto}.
The proposed BSL implementation is fully compatible with standard HMC and straightforward to deploy in a probabilistic framework such as Stan.
}
Third, we outline how BSL can be applied to enhance ML-NMR (in Section~\ref{sec:sec4}), and using the network of psoriasis clinical trials dataset from \citet{phillippo2020multilevel}, we demonstrate that BSL-enhanced ML-NMR can substantially improve parameter estimation compared to standard ML-NMR, recovering much of the information lost when individual covariate values are missing (in Section~\ref{sec:sec5}). We conclude in Section~\ref{sec:sec6} with a discussion of limitations, connections to related work on synthetic data generation, and directions for future research including extensions to time-to-event outcomes and alternative likelihood-free inference strategies.


\section{BSL for Missing Data: A Simple Example}
\label{sec:sec2}

We begin with a simple example that illustrates how BSL can incorporate ancillary summary information when individual-level data are only partially observed. The setting is deliberately minimal so that the core BSL mechanism is transparent before we apply it to the full ML-NMR problem in Section~\ref{sec:sec4}.

\subsection{Setup}

Consider $n$ independent and identically distributed normal observations
with unknown mean $\mu$ and known standard deviation $\sigma$:
\begin{equation}
Y_i \;\overset{\text{iid}}{\sim}\; \text{Normal}(\mu,\, \sigma^2),
\quad i = 1, \ldots, n.
\end{equation}
Suppose we observe only the first $m < n$ values, $y_{1:m}$, but
additionally have access to a summary statistic computed from the
\textit{complete} data:
\begin{equation}
S \;=\; \frac{1}{n} \sum_{i=1}^{n} \mathbb{I}(Y_i > c),
\label{eq:summary_stat}
\end{equation}
the proportion of all $n$ observations exceeding a known threshold $c$.
This mirrors the situation described in the introduction: individual-level
data are partially withheld, but an aggregate summary derived from the
full dataset is available.

\subsection{The inferential challenge}

With both the observed data and the summary statistic in hand, the
target posterior is
\begin{align}
\pi(\mu \mid y_{1:m},\, s)
&\;\propto\; f(y_{1:m},\, s \mid \mu)\, \pi(\mu) \nonumber \\
&\;=\; f(s \mid y_{1:m},\, \mu)\;
       f_{\text{Norm}}(y_{1:m} \mid \mu)\;
       \pi(\mu),
\label{eq:target_posterior}
\end{align}
where $\pi(\mu)$ is the prior, and $f_{\text{Norm}}(y_{1:m} \mid \mu)$ is the normal density function for the observed data. The likelihood of the observed data is straightforward. The difficulty lies in $f(s \mid y_{1:m},\, \mu)$: the conditional density of the summary statistic given the observed data and the parameter. For the indicator-based summary in~\eqref{eq:summary_stat}, this density depends on the distribution of the unobserved values $Y_{(m+1):n}$ and is not available in closed form for general thresholds and parameter values.

\subsection{BSL approximation}
\label{sec:bsl_approx}

BSL sidesteps the intractability of $f(s \mid y_{1:m},\, \mu)$ by
approximating it through simulation. The key idea is that while we
cannot evaluate this density analytically, we \textit{can} generate
synthetic realizations of $S$ by imputing the missing data under the
current parameter value, and then use the empirical distribution of
these synthetic summaries as a surrogate likelihood.

At each MCMC iteration with current parameter value $\mu$ (and known $\sigma$), BSL proceeds in three steps:

\begin{enumerate}
\item \textbf{Impute the missing data.} Generate $B$ synthetic
  completions of the unobserved values:
  \begin{equation}
  y_{(m+1):n}^{(b)} \;\sim\;
  \text{Normal}(\mu,\, \sigma^2),
  \quad b = 1, \ldots, B.
  \end{equation}

\item \textbf{Compute synthetic summaries.} For each synthetic
  completion, evaluate the summary statistic:
  \begin{equation}
  \hat{s}_b \;=\; \frac{1}{n} \left(
    \sum_{i=1}^{m} \mathbb{I}(y_i > c) \;+\;
    \sum_{i=m+1}^{n} \mathbb{I}\!\left(y_i^{(b)} > c\right)
  \right),
  \quad b = 1, \ldots, B.
  \end{equation}

\item \textbf{Form the synthetic likelihood.} Approximate the
  conditional density with a normal:
  \begin{equation} \label{eq:synthetic-likelihood}
  f(s \mid y_{1:m},\, \mu)
  \;\approx\;
  f_{\text{Norm}}\!\left(s \;\big|\; \hat{\mu}_S,\, \hat{\sigma}_S^2\right),
  \end{equation}
  where $\hat{\mu}_S$ and $\hat{\sigma}_S^2$ are the sample mean and
  variance of $\{\hat{s}_b\}_{b=1}^B$.
\end{enumerate}

\noindent The approximate posterior is then
\begin{equation}
\pi(\mu \mid y_{1:m},\, s)
\;\approx\;
f_{\text{Norm}}\!\left(s \;\big|\; \hat{\mu}_S,\,
  \hat{\sigma}_S^2\right)\;
f_{\text{Norm}}(y_{1:m} \mid \mu)\;
\pi(\mu).
\label{eq:bsl_posterior}
\end{equation}
Note that $\hat{\mu}_S$ and $\hat{\sigma}_S^2$ are both functions of
$\mu$ (through the imputed data), so they are recomputed at every MCMC
iteration. An important feature of BSL is that it requires no tolerance
or bandwidth hyperparameters: the empirical variance of the synthetic
summaries naturally calibrates the width of the surrogate likelihood.

Also note that if the summary statistic is multivariate (i.e.,
multiple summaries are available), the univariate normal approximation
generalizes to a multivariate normal,
$\text{MVN}(\hat{\boldsymbol{\mu}}_S,\, \hat{\boldsymbol{\Sigma}}_S)$,
where $\hat{\boldsymbol{\mu}}_S$ and
$\hat{\boldsymbol{\Sigma}}_S$ are the sample mean vector and covariance
matrix of the synthetic summary vectors.

\subsection{Numerical illustration}
\label{sec:numerical_illustration}

We illustrate the BSL approach with simulated data. We generate $n =
120$ observations from $\text{Normal}(\mu = 1,\, \sigma = 1)$, of which
only $m = 10$ are observed. The summary statistic is the proportion of
all $n$ observations exceeding the threshold $c = 2$.

Table~\ref{tab:results_example1} presents posterior summaries for $\mu$
under three approaches: the \textit{oracle} posterior using all $n$
observations, the \textit{simple} posterior using only the $m$ observed
values, and \textit{BSL} posterior using the $m$ observed values plus the
summary statistic, $s$.  We use $B = 25$ synthetic replicates for the BSL. All three approaches are implemented in Stan; the details of how BSL is implemented within Stan's framework will be reviewed in Section~\ref{sec:sec3}.

\begin{table}[htbp]
\centering
\begin{tabular}{lcccccc}
\hline
Method & Mean & SD & 95\% CrI & CrI Ratio & Time (s) & $\hat{R}$ \\
\hline
Oracle (complete data) & 1.21 & 0.09 & (1.04, 1.39) & 1.00 & 2.8 & 1.000 \\
  Simple (observed only) & 0.86 & 0.32 & (0.26, 1.50) & 3.51 & 3.1 & 1.000 \\
  BSL & 1.23 & 0.12 & (1.00, 1.46) & 1.28 & 131.5 & 1.002 \\
   \hline
\end{tabular}
\caption{Posterior inference for $\mu$ with $\sigma$ known ($B = 25$
    synthetic replicates). CrI Ratio is the width of the 95\% credible
    interval relative to the oracle (complete data) posterior; values
    near 1 indicate that the method recovers most of the information
    in the complete data.}
\label{tab:results_example1}
\end{table}

Note that the CrI ratio reports the width of each method's 95\% credible interval (CrI) relative to the oracle.  The simple posterior, based on only 10 observations, produces a wide CrI (CrI ratio of 3.51) and a point estimate that is pulled away from the value obtained by the oracle (i.e., the value one would estimate in the absence of any missing data). The BSL posterior, which additionally incorporates the exceedance proportion, recovers a CrI ratio of 1.28 with a point estimate that closely matches the oracle analysis. This demonstrates that even a single summary statistic can
recover a substantial amount of the information lost to missingness,
provided that the summary is informative about the parameter of interest.


\section{Implementing BSL with HMC (in Stan)}
\label{sec:sec3}

In this section, we detail the computational challenges and proposed solutions for implementing our the BSL procedure with HMC.  Readers who are less interested in the computational/MCMC aspects may wish to skip to Section ~\ref{sec:sec4}.     

The BSL procedure described in Section~\ref{sec:sec2} requires three
capabilities at each MCMC iteration: generating synthetic datasets by
imputing missing values, computing summary statistics from the imputed
data, and evaluating a synthetic likelihood that compares these summaries
to the observed ones.
These requirements are not immediately compatible with popular implementation of HMC, such as the one found in Stan, and additional steps must be taken to obtain an implementation which is both computationally efficient and statistically accurate.


\subsection{HMC and Coding Blocks in Stan}

\textcolor{black}{
To understand the constraints imposed by HMC, we briefly review certain mechanisms that underlie the algorithm; see 
\citet{neal2012mcmc} and \citet{betancourt2017hmc} for a more in-depth introduction. We focus on the implementation in Stan, even though our discussion generalizes to other probabilistic frameworks.
}

\textcolor{black}{
At each iteration of MCMC, HMC generates a Hamiltonian trajectory driven by the gradient of the log posterior.
In all but the simplest cases, this trajectory cannot be calculated exactly and instead a discretized trajectory is computed via a numerical integrator.
The Hamiltonian trajectory is then obtained by taking a certain number of integration steps, with each step requiring a gradient evaluation.
The next MCMC draw is then sampled from the trajectory, according to a scheme that corrects for error in the numerical integrator and preserves detailed balance.
}

\textcolor{black}{
Stan is a language to specify a Bayesian model, that is a program to calculate the log joint density, typically decomposed into a log prior and a log likelihood.
(The gradient is calculated automatically via automatic differentiation \citep{baydin2018automatic, margossian2019review}.)
Calculation and differentiation of the log joint density occurs in two coding blocks of Stan: the \texttt{transformed parameters} and the \texttt{model} block.
For our discussion, the distinction between the two blocks is immaterial and so we will refer to them jointly as the \texttt{model} blocks.
Stan also provides a \texttt{generated quantities} block, which allows for post-hoc calculations which depend on the model parameters but do not contribute to the log joint density.
An example of such calculations would be model predictions.
Operations in the \texttt{generated quantities} block are cheap: they occur once per MCMC iteration.
By contrast, operations in the \texttt{model} blocks occur several times per iterations---once per integration step when simulating a Hamiltonian trajectory---and need to be differentiated.
Hence, computation is almost entirely dominated by operations in the \texttt{model} blocks.
Table~\ref{tab:blocks} provides a summary of the block structure in Stan.}

\textcolor{black}{
The details of HMC's implementation manifest as (soft) constraints in Stan.
First, Stan's HMC expects the log likelihood to be differentiable.
Second, Stan's HMC requires exact calculations of the log joint distribution and its gradient, rather than a stochastic estimation thereof (as would be obtained with a synthetic likelihood via eq.~\eqref{eq:synthetic-likelihood}).
As a result, operations in the \texttt{model} blocks must be differentiable and cannot entail any random element.
On the other hand, there are no such constraints for operations in the \texttt{generated quantities} block, since these do not contribute to the evaluation of the log joint density.
}

\begin{table}
  \begin{center}
    \begin{tabular}{l l l}
    \rowcolor{LightCyan}
    {\bf Coding block} & {\bf Number of operations} & {\bf Requirements} \\
    \texttt{transformed parameters} & Multiple evaluations and differentiation & Deterministic, \\
    \texttt{+ model} &  per MCMC iteration & differentiable \\
    \rowcolor{LightCyan} \texttt{generated quantities} & One evaluation per MCMC iteration. & None \\
    \end{tabular}
  \end{center}
  \caption{Coding blocks in Stan. The \texttt{transformed parameters}
  and \texttt{model} blocks are used to evaluated the log joint density.
  The \texttt{generated quantities} block is used for post-sampling calculations which are relatively cheap.
  }
  \label{tab:blocks}
\end{table}

\subsection{Common random numbers}
\label{sec:crn}

BSL requires random number generation to produce synthetic datasets at
each MCMC iteration. Stan, however, does not permit random number
generation within the \texttt{model} blocks because it expects the log likelihood to be a deterministic function of the parameters and the data.
\textcolor{black}{
The use of stochastic estimators of the log likelihood and its gradient for HMC has been explored, notably for subsampling, a strategy used in large data problems where the likelihood is evaluated using only a random subsample of the data at each iteration \citep{neiswanger2013asymptotically}.
In this context, it was found that the combination of HMC and subsampling can be inefficient \citep{betancourt2015incompatibility, chen2014sghmc} and these analyses raise questions about the effectiveness of HMC with stochastic estimators in general.
 Still, if the gradient is estimated with high accuracy, the error in HMC can be controlled.
For BSL, we require $B$ to be sufficiently large and we will see in Section~\ref{sec:psis} how to check that this is indeed the case.}

\textcolor{black}{To engineer a synthetic likelihood in Stan,} we generate all
required random numbers \textit{before} sampling begins and pass them into Stan as fixed ``data''. The synthetic data generation then becomes a deterministic, differentiable transformation of these pre-drawn values and the current parameters. This is sometimes known as ``shipping in'' randomness from outside of the MCMC \citep{ship}.

In the simple example of Section~\ref{sec:sec2} we can ``ship in'' the required randomness by recognizing that generating a synthetic
missing observation $y_i^{(b)} \sim \text{Normal}(\mu, \sigma^2)$ is
equivalent to computing
\begin{equation}
y_i^{(b)} = \mu + \sigma \, z_i^{(b)},
\quad \text{where} \quad z_i^{(b)} \sim \text{Normal}(0, 1).
\label{eq:crn_normal}
\end{equation}
Then, the values $z_i^{(b)}$ can be drawn once before the MCMC run begins and stored as a matrix $\mathbf{Z} \in \mathbb{R}^{B \times (n-m)}$, which is passed to Stan as data. In the \texttt{model} blocks, the synthetic observations are computed as a deterministic function of $\mu$, $\sigma$, and $\mathbf{Z}$. The same random numbers are reused across all MCMC iterations, hence the term ``common random numbers.''

This approach has two important consequences. First, the synthetic data
generation is fully deterministic conditional on $\mathbf{Z}$, so it
does not violate Stan's requirement that the target log-density be a
deterministic function of the parameters. Second, because the same
$\mathbf{Z}$ values are used at every iteration, the synthetic
summaries vary smoothly as $\mu$ changes, which can, in theory, improve the
stability of the Hamiltonian dynamics
\citep{meeds2015hamiltonian}.

\subsection{Sufficient statistic representation}
\label{sec:sufficient_stat}

The BSL implementation from Section~\ref{sec:bsl_approx} generates $(n - m)$ synthetic values per
replication, requiring $O(B \times (n-m))$ operations per MCMC
iteration. For large $n - m$ this can be expensive, and the cost will no doubt grow
further in more complex models where the missing data are high-dimensional.

In many cases, the summary statistic depends on the missing data only through a lower-dimensional sufficient quantity. In our simple example, the exceedance proportion depends on the $(n - m)$ missing values only
through the count of how many exceed the threshold~$c$. Given the current parameter value $\mu$ (with $\sigma$ known), the probability that a single observation exceeds $c$ is
\begin{equation}
p^* = 1 - \Phi\!\left(\frac{c - \mu}{\sigma}\right),
\end{equation}
where $\Phi(\cdot)$ is the standard normal CDF. Then, recognizing that the number of missing observations exceeding $c$ follows a binomial
distribution, we have:
\begin{equation}
X^{(b)} \sim \text{Binomial}(n - m,\; p^*),
\quad b = 1, \ldots, B.
\end{equation}
Generating the synthetic data in this way will reduce the per-iteration cost from $O(B \times (n-m))$ to $O(B)$.

\subsection{Continuous relaxation}
\label{sec:continuous_relaxation}


\textcolor{black}{The next challenge we must address concerns the requirement that the log likelihood is differentiable with respect to the model parameters.
In this respect,}
discrete elements in the synthetic data generation are problematic regardless of
whether a sufficient statistic representation is used. The indicator
function $\mathbb{I}(y_i^{(b)} > c)$ in the individual-level BSL of
Section~\ref{sec:bsl_approx} already introduces discontinuities, and
the binomial reformulation retains this issue: $X^{(b)}$ takes integer
values, so the synthetic summaries $\hat{s}_b$ change in discrete steps
as $\mu$ varies. 

\textcolor{black}{
Several variants of HMC have been proposed to target non-smooth distributions \citep[e.g.,][]{betancourt2010nested, pakman2013auxiliary, pakman2014exact, afshar2015reflection, zhou2019lfppl, nishimura2020discontinuous}.
These methods often require specialized constructions which are attentive to the particular structure of the discontinuity, although recent algorithms \citep{nishimura2020discontinuous} are more general.
Investigating the role of non-smooth HMC for BLS is left to future work; in this paper, we focus on standard HMC.
}

\textcolor{black}{
In standard HMC, the numerical integrator used to simulate a Hamiltonian trajectory ignores changes in the log likelihood due to discontinuities, because the integrator only has access to information from the gradient.
Hence, the simulated Hamiltonian trajectory is inaccurate for discontinuous targets.
The final step of HMC---weighted sampling across the trajectory---corrects for inaccuracy in the simulated trajectory and so the Markov chain still has the correct stationary distribution.
But without the exploration afforded by (accurate) Hamiltonian dynamics, HMC becomes inefficient.
The problem is exacerbated in Stan, because adaptive HMC reacts to the inaccuracy of the Hamiltonian trajectory by reducing the step size of the numerical integrator.
Not only does this strategy \textit{not} fix the problem, it increases the number of gradient evaluations required to simulate a trajectory of a given length.
The end result is poor mixing per iteration, with the computational cost of each iteration increased.
}

\textcolor{black}{The pathology described above does not systematically manifest.
We can get away with some amount of discontinuity, if the sampler generates few Hamiltonian trajectories that cross a discontinuity line.}
In the simple one-parameter example of Section~\ref{sec:sec2}, the HMC sampler
was able to mix adequately despite this discreteness. In higher
dimensions, however, we observe that discrete jumps in the synthetic likelihood surface cause mixing to deteriorate rapidly. For more complex applications, BSL implementation in Stan therefore requires a way to smooth any discrete elements in the synthetic likelihood construction.

A general approach to smooth discrete distributions in gradient-based frameworks is to do a ``continuous relaxation,'' \textcolor{black}{an approach which has worked well in certain specific cases \citep[e.g.,][]{zhang2012continuous, pakman2013auxiliary, grathwohl2021oops}}. 
In our proposed model, we replace a discrete distribution with a continuous approximation. For the binomial case in our simple example, we replace
\begin{equation}
X^{(b)} \sim \text{Binomial}(n - m,\; p^*)
\end{equation}
with the normal approximation
\begin{equation}
X^{(b)} \sim \text{Normal}\Big((n - m)\, p^*,\; \sqrt{(n - m)\, p^* (1 - p^*)}\Big)
\end{equation}
or written alternatively as:
\begin{equation}
\tilde{X}^{(b)} = (n - m)\, p^* +
  \sqrt{(n - m)\, p^* (1 - p^*)}\; w^{(b)},
\label{eq:continuous_relaxation}
\end{equation}
where $w^{(b)} \sim \text{Normal}(0, 1)$ is a pre-drawn standard normal
(following the common random numbers strategy of Section~\ref{sec:crn}). The result $\tilde{X}^{(b)}$ is clamped to the
interval $[0,\, n - m]$ to remain in the valid range. 
 (In practice, the clamping only occurs when $p^*$
 is extreme, so the resulting non-differentiability due to clamping should have only minor impact.)
The synthetic summary statistic then becomes
\begin{equation}
\hat{s}_b = \frac{1}{n}\left(
  \sum_{i=1}^{m} \mathbb{I}(y_i > c) + \tilde{X}^{(b)}
\right).
\end{equation}
Crucially, $\tilde{X}^{(b)}$ is now a smooth, differentiable function
of $\mu$ (through $p^*$), and the pre-drawn values $w^{(b)}$ are fixed
data. The entire synthetic data generation pipeline -- from parameters
to synthetic summaries to synthetic likelihood -- is therefore
differentiable, and can be placed in Stan's \texttt{model} blocks.
\textcolor{black}{On the other hand, the continuous approximation alters the stationary distribution of HMC. 
We will address this bias in Section~\ref{sec:psis}.
}

For the ML-NMR setting which we will discuss in Section~\ref{sec:sec4}, the discrete
distribution to be relaxed is a multinomial rather than a binomial.  We will handle this by applying the normal approximation sequentially:
for each category $k = 1, \ldots, K-1$, we draw from a normal
approximation to the conditional binomial given the remaining count
and remaining probability mass, clamping to $[0, n_{\text{remaining}}]$
and subtracting the assigned count before proceeding to the next category. The final category $K$ receives the remainder. This sequential conditional approach preserves the constraint that the counts sum to the marginal total while maintaining differentiability
at each step.

\subsection{Importance sampling correction}
\label{sec:psis}

While the continuous relaxation strategy improves HMC mixing, it also introduces bias.
We propose correcting for this discrepancy with a post-hoc importance sampling step \citep{vehtari2024pareto}.

Let $\ell_{\text{cont}}(\theta)$ denote the
log synthetic likelihood evaluated using the continuous relaxation
(i.e., the quantity computed in the \texttt{model} block),
and let $\ell_{\text{disc}}(\theta)$ denote the log synthetic likelihood
that would be obtained from exact discrete sampling. The MCMC samples
are drawn from a posterior 
of the form:
\begin{equation}
\pi_{\text{cont}}(\theta \mid y_{1:m}, s)
\;=\;
\frac{1}{Z_\text{cont}} \ \exp\!\big(\ell_{\text{cont}}(\theta)\big)\;
f_{\text{Norm}}(y_{1:m} \mid \theta)\;
\pi(\theta),
\end{equation}
where $Z_\text{cont}$ is the unknown normalizing constant,
but we wish to target
\begin{equation}
\pi_{\text{disc}}(\theta \mid y_{1:m}, s)
\;=\; \frac{1}{Z_\text{disc}} \ 
\exp\!\big(\ell_{\text{disc}}(\theta)\big)\;
f_{\text{Norm}}(y_{1:m} \mid \theta)\;
\pi(\theta).
\end{equation}
The importance weights are
\begin{equation}
w(\theta) = \frac{Z_\text{disc}}{Z_\text{cont}} \  \exp\!\big(\ell_{\text{disc}}(\theta) -
  \ell_{\text{cont}}(\theta)\big).
\label{eq:is_weights}
\end{equation}
\textcolor{black}{
In practice, we cannot compute the importance weights exactly because of the unknown normalizing constants, however we can construct an importance sampling estimator with self-normalized weights.
}

Computing $\ell_{\text{disc}}(\theta)$ requires drawing from the exact
discrete distribution (binomial or multinomial), which involves random
number generation. 
This can be done in Stan's \texttt{generated quantities} block, which executes after each MCMC draw
and permits the use of Stan's \texttt{\_rng} functions
\textcolor{black}{(or it can be done outside of Stan entirely)}. At each
iteration, we draw a large number of discrete synthetic datasets (e.g., using \texttt{binomial\_rng} or \texttt{multinomial\_rng}), compute the
corresponding synthetic summaries and their mean and covariance, and
evaluate the discrete synthetic likelihood. The \textcolor{black}{(unnormalized)} log importance weight
$\ell_{\text{disc}}(\theta) - \ell_{\text{cont}}(\theta)$ is then stored as
a generated quantity and used for post-hoc reweighting.

In practice, we apply PSIS \citep{vehtari2024pareto} to stabilize the weights. PSIS fits a generalized Pareto distribution to the upper tail of the importance
weights and replaces extreme values with order-statistic-based smoothed values. The estimated shape parameter $\hat{k}$ of the Pareto tail provides a diagnostic: $\hat{k} < 0.7$ suggests reliable importance sampling, while larger values indicate that the continuous approximation may be too far from the discrete target.

Note that by using a much larger number of synthetic replicates for PSIS in the \texttt{generated quantities} block (e.g., $B_{\text{disc}} = 1000$) than for posterior sampling in the \texttt{model} blocks (e.g., $B = 25$), the discrete synthetic likelihood is estimated with negligible Monte Carlo error. The $\hat{k}$ diagnostic therefore reflects not only the quality of the continuous relaxation but also the adequacy of the finite-$B$ estimation of $\hat{\mu}_S$ and $\hat{\Sigma}_S$ used during sampling. If one obtains a large value of $\hat{k}$, a larger number
of synthetic replicates $B$ and/or a better continuous approximation strategy may be required.

\subsection{Summary of the implementation strategy}
\label{sec:implementation_summary}

Combining the four ideas above, we propose the complete BSL implementation in Stan
proceeds as follows:

\begin{enumerate}
\item \textbf{Before sampling:} Generate all required random numbers
  (e.g., standard normals $w^{(b)}$ for the continuous relaxation) and pass
  them to Stan as data, along with the observed data and summary
  statistics.

\item \textbf{Transformed parameters block:} Using the current parameter
  values $\theta$ and the pre-drawn random numbers, compute the
  model-implied summary values (e.g., probability $p^*$), generate continuous synthetic data (e.g., via the normal approximation), compute synthetic summary statistics, and compute the synthetic mean $\hat{\mu}_S$ and covariance $\hat{\Sigma}_S$.

\item \textbf{Model block:} Add the usual data likelihood and priors to
  the target, plus the synthetic likelihood:
  \begin{equation}
  \text{target} \mathrel{+}=
    \log f_{\text{MVN}}\!\left(\mathbf{s}_{\text{obs}} \;\big|\;
    \hat{\boldsymbol{\mu}}_S,\;
    \hat{\boldsymbol{\Sigma}}_S\right).
  \end{equation}

\item \textbf{Generated quantities block:} Draw fresh discrete samples
  using Stan's RNG functions, compute the discrete synthetic likelihood
  $\ell_{\text{disc}}(\theta)$, and output both
  $\ell_{\text{disc}}(\theta)$ and $\ell_{\text{cont}}(\theta)$ for
  post-hoc PSIS correction.
\end{enumerate}

\noindent This strategy keeps the \texttt{model} block expressions fully
differentiable while providing a principled correction for the
approximation error via importance sampling. The PSIS diagnostic
$\hat{k}$ provides an automatic check on the quality of the continuous
relaxation and can help determine if the $B$ value is large enough.

\subsection{Numerical illustration (continued)}
\label{sec:numerical_illustration_sec3}

We return to the simple example of Section~\ref{sec:numerical_illustration}
to demonstrate the implementation strategies described above. We now
add two methods to the comparison: BSL with continuous relaxation of the
binomial (Section~\ref{sec:continuous_relaxation}), and the same method
with PSIS correction (Section~\ref{sec:psis}). All settings ($n = 120$,
$m = 10$, $c = 2$, $B = 25$) are as before.

\begin{figure}[htbp]
    \centering
    \includegraphics[width=0.90\linewidth]{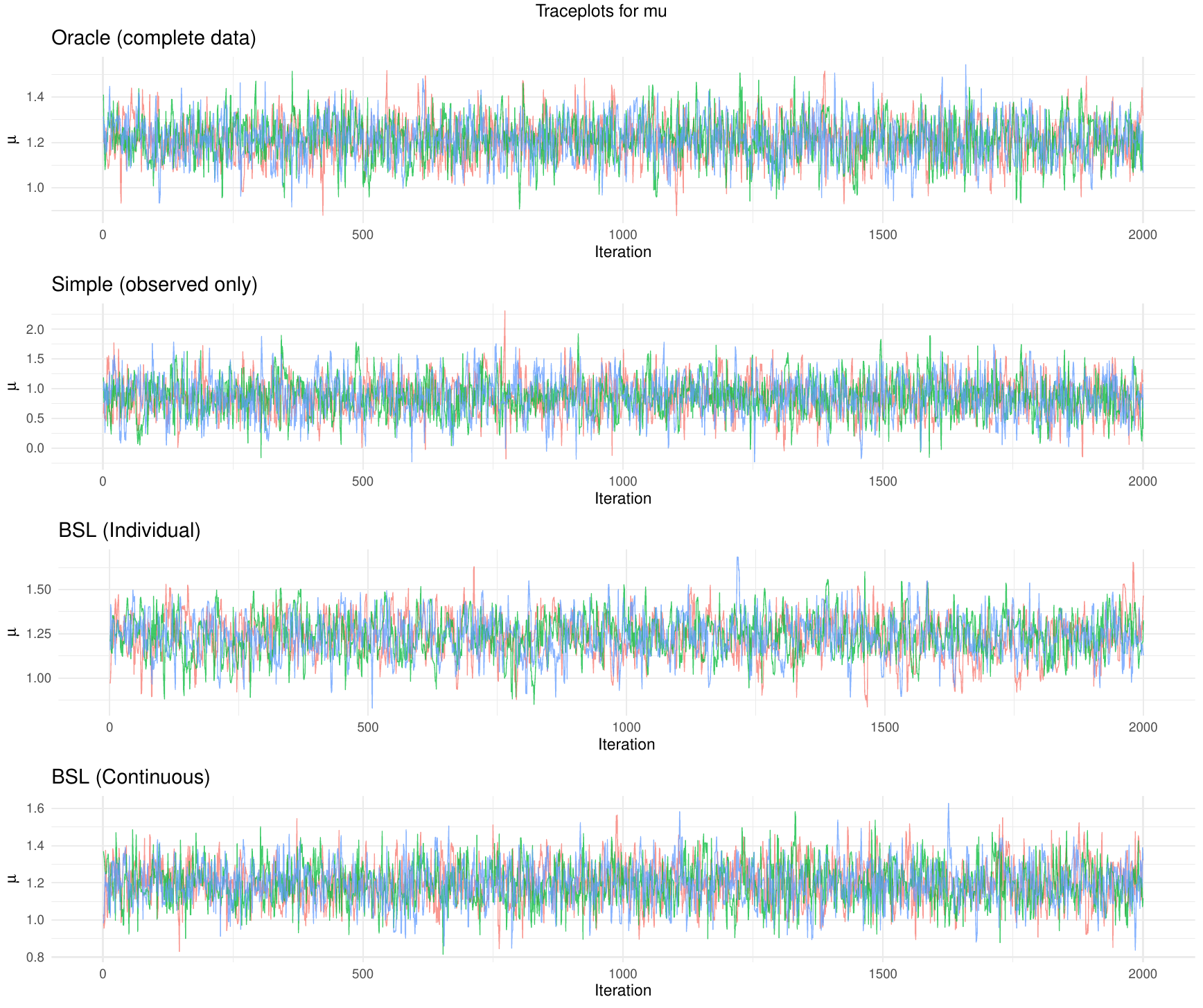}
    \caption{Trace-plots of $\mu$ for each method, with chains shown in different colors for the ``simple example''.}
    \label{fig:traceplot_example1}
\end{figure}

Table~\ref{tab:results_example1_sec3} extends the results from
Table~\ref{tab:results_example1}; Figure \ref{fig:traceplot_example1} shows trace-plots. The continuously relaxed BSL variant
produces posterior summaries comparable to the individual-level BSL from
Section~\ref{sec:sec2}, confirming that the normal approximation to the
binomial preserves inferential quality in this setting. The PSIS
correction has minimal effect here, with a Pareto $\hat{k}$ well below
0.7, indicating that the continuous relaxation is a good approximation
to the discrete distribution for this problem.

\begin{table}[htbp]
\centering
\begin{tabular}{lccccccc}
\hline
Method & Mean & SD & 95\% CrI & CrI Ratio & Time (s) & $\hat{R}$ & $\hat{k}$ \\
\hline
Oracle (complete data) & 1.21 & 0.09 & (1.04, 1.39) & 1.00 & 2.8 & 1.000 & -- \\
  Simple (observed only) & 0.86 & 0.32 & (0.26, 1.50) & 3.51 & 3.1 & 1.000 & -- \\
  BSL (Individual) & 1.23 & 0.12 & (1.00, 1.46) & 1.28 & 131.5 & 1.002 & -- \\
  BSL (Continuous) & 1.20 & 0.11 & (0.98, 1.41) & 1.21 & 26.0 & 1.000 & -- \\
  BSL (Continuous + PSIS) & 1.22 & 0.12 & (1.00, 1.45) & 1.28 & 26.0 & 1.000 & 0.203 \\
   \hline
\end{tabular}
\caption{Posterior inference for $\mu$ with $\sigma$ known ($B = 25$
    synthetic replicates, $B_{\text{disc}} = 1000$ for PSIS correction).
    CrI Ratio is the width of the 95\% credible interval relative to the
    oracle (complete data) posterior. $\hat{k}$ is the Pareto shape
    diagnostic for importance sampling (values below 0.7 indicate reliable
    correction).}
\label{tab:results_example1_sec3}
\end{table}

The key advantage of the continuously relaxed implementation is not visible in this simple one-parameter example, but it becomes critical in more complex models.  In the ML-NMR application of Section~\ref{sec:sec4}, where the parameter space includes study intercepts, treatment effects, prognostic effects, and effect modifier interactions, this strategy will be essential for tractable inference.


\section{BSL-Enhanced ML-NMR}
\label{sec:sec4}

We now apply the BSL framework of Sections~\ref{sec:sec2}
and~\ref{sec:sec3} to multilevel network meta-regression (ML-NMR)
\citep{phillippo2020multilevel}. 



\subsection{Data structure}

Consider a network of $J$ randomized studies comparing $T$ treatments for a binary outcome. For individual $i$ in study $j$ receiving treatment $\text{trt}_{ij} \in \{1, \ldots, T\}$, we observe outcome $y_{ij} \in \{0,1\}$ and potentially a vector of baseline covariates $x_{ij} \in \mathbb{R}^P$. The data structure varies across three types of studies, defined based on data-availability:

\begin{itemize}
\item \textbf{Full IPD studies} ($j \in \mathcal{J}_{\text{full}}$): We observe $(y_{ij}, \text{trt}_{ij}, x_{ij})$ for all individuals.
\item \textbf{Partial IPD studies with subgroup results} ($j \in \mathcal{J}_{\text{BSL}}$): We observe $(y_{ij}, \text{trt}_{ij})$ but individual covariates $x_{ij}$ are unavailable. However, the study reports:
\begin{enumerate}
\item the marginal covariate distribution $\pi_j$ for the study sample (or enough information with which to infer this distribution and assume it is known), and
\item subgroup-specific treatment effects $s_{j}$ (e.g., log-ORs stratified by covariate levels).
\end{enumerate}
\item \textbf{Partial IPD studies without subgroup results} ($j \in \mathcal{J}_{\text{partial}}$): We observe $(y_{ij}, \text{trt}_{ij})$ and $\pi_j$, but no subgroup summaries are available.
\end{itemize}

\noindent Denote by $n_j$ the number of individuals in study $j$ and by $n_{j,t}$ the number receiving treatment $t$ in study $j$. For studies in $\mathcal{J}_{\text{BSL}}$, let $s_j \in \mathbb{R}^{D_j}$ denote the vector of $D_j$ reported subgroup summaries, where $D_j$ may vary across studies depending on what is reported.

The key challenge is that standard ML-NMR uses only $(y_{ij}, \text{trt}_{ij}, \pi_j)$ for the partial IPD studies, effectively discarding any valuable information in $s_j$. We propose to incorporate $s_j$ using  BSL.

\subsection{Model specification}
Let $Y_{ij}$ be a binary random variable indicating response for individual $i$ in study $j$.  Then we model the probability of response for individual $i$ in study $j$ as:
\begin{align}
(Y_{ij} \mid \text{trt}_{ij}, x_{ij})
  &\sim \text{Bernoulli}(\theta_{ij}) \nonumber \\
\text{logit}(\theta_{ij})
  &= \mu_j + \gamma_{\text{trt}_{ij}}
     + x_{ij}^\top (\beta_1 + \beta_{2,\text{trt}_{ij}}),
\label{eq:logit_model}
\end{align}
where $\mu_j$ is the study-specific baseline log-odds,
$\gamma_t$ is the treatment effect for treatment $t$ relative to the
reference (treatment~1), $\beta_1 \in \mathbb{R}^P$ captures prognostic
covariate effects, and $\beta_{2,t} \in \mathbb{R}^P$ captures
treatment-by-covariate interactions (effect modification) for
treatment~$t$.   Let $\theta = (\mu_{1:J},\, \gamma_{1:T},\, \beta_1,\, \beta_{2,1:T})$.  Note that the logit link is a particular choice and the outcome model could use a different link function if required.  For identifiability, we set $\gamma_1 = 0$ and $\beta_{2,1} = \mathbf{0}$, so that $\beta_{2,t}$ represents how the effect of treatment $t$ (relative to treatment~1) varies with
covariates.

In some applications, one may assume that effect modification is shared across treatments (or across all treatments within a given class), i.e., $\beta_{2,t} = \beta_2$ for all $t > 1$. This \textit{shared effect modifier} assumption reduces the number of parameters and can often be necessary for estimation when treatment-specific interactions cannot be identified from sparse data \citep{phillippo2023validating}.

Covariates $x_{ij}$ may be continuous or binary; the model
specification~\eqref{eq:logit_model} accommodates both. For full IPD
studies, the actual covariate vectors enter the likelihood directly. For
partial IPD and aggregate studies, where individual covariate values are
unavailable, the likelihood must be marginalized over the covariate
distribution. Following \citet{phillippo2020multilevel}, we approximate
this marginalization using numerical integration over a finite set of
$K$ representative covariate patterns. Specifically, for each study $j$
lacking individual covariates, we construct $K$ integration points
$Z_{j,1}, \ldots, Z_{j,K} \in \mathbb{R}^P$ with equal weights
$\pi_{jk} = 1/K$. These integration points are study-specific, drawn
from each study's (assumed) known covariate distribution to capture between-study heterogeneity in covariate profiles (e.g., in \citet{phillippo2023r}'s psoriasis  example, there are $K = 64$ integration points determined via  Quasi-Monte Carlo numerical integration using a Gaussian copula and Sobol' sequences \citep{phillippo2020multilevel}).  We denote by $\eta_{j,t,k} = \mu_j + \gamma_t + Z_{j,k}^\top(\beta_1 + \beta_{2,t})$ the linear predictor evaluated at integration point $k$.

\subsection{Likelihood components}

The full likelihood combines contributions from each study type. For
studies $j \in \mathcal{J}_{\text{full}}$ with complete individual-level
data, the likelihood uses the actual covariate vectors:
\begin{equation}
\mathcal{L}_{\text{full}}(\theta) = \prod_{j \in \mathcal{J}_{\text{full}}} \prod_{i} \text{Bernoulli-Logit}(y_{ij} \mid \eta_{ij}),
\end{equation}
where $\eta_{ij} = \mu_j + \gamma_{\text{trt}_{ij}} +
x_{ij}^\top(\beta_1 + \beta_{2,\text{trt}_{ij}})$.

For studies $j \in \mathcal{J}_{\text{partial}} \cup
\mathcal{J}_{\text{BSL}}$ where individual covariates are missing,
ML-NMR marginalizes over the $K$ integration points with equal weights.
Aggregating by $(y, t)$ strata:
\begin{equation}
\mathcal{L}_{\text{partial}}(\theta) = \prod_{j} \prod_{y,t} \left( \frac{1}{K}\sum_{k=1}^{K}  \text{Bernoulli-Logit}(y \mid \eta_{j,t,k}) \right)^{n_{j,y,t}},
\label{eq:marginal_likelihood}
\end{equation}
where $n_{j,y,t}$ is the count of individuals in study $j$ with outcome
$y$ and treatment $t$. This marginalization avoids aggregation bias
while remaining computationally tractable.

For studies $j \in \mathcal{J}_{\text{BSL}}$, we incorporate the
subgroup summary statistics $s_j$ via synthetic likelihood:
\begin{equation}
f(s_j \mid y_j, \text{trt}_j, \theta) \approx f_{\text{MVN}}(s_j \mid \hat{\mu}_{S,j}, \hat{\Sigma}_{S,j}),
\end{equation}
where $\hat{\mu}_{S,j}$ and $\hat{\Sigma}_{S,j}$ are the mean vector
and covariance matrix of synthetic summary statistics computed via the
following BSL procedure. At each MCMC iteration with current parameter
value $\theta$, we approximate the synthetic likelihood for each study
$j \in \mathcal{J}_{\text{BSL}}$ as follows:

\begin{enumerate}
\item \textbf{Compute conditional pattern probabilities.} For each
  combination of outcome $y \in \{0,1\}$ and treatment $t \in \{1,
  \ldots, T\}$ available in study $j$, compute the probability that an
  individual with outcome $y$ and treatment $t$ belongs to integration
  point $k$ (i.e., has covariate values approximately matched by those represented at integration point $k$). Let $C \in \{1,\ldots,K\}$ denote the integration-point index
  (covariate pattern) indicating which study-specific integration point $Z_{j,k}$  an individual belongs to, with a uniform prior $\Pr(C=k)=\pi_{jk}=1/K$. Then
\begin{equation}
\textrm{Pr}(C = k \mid y, t, \theta) = \frac{ \textrm{Pr}(Y = y \mid C = k, \text{Trt} = t, \theta)}{\sum_{k'=1}^K  \textrm{Pr}(Y = y \mid C = k', \text{Trt} = t, \theta)},
\label{eq:posterior_pattern_prob}
\end{equation}
where the equal prior weights $1/K$ cancel in the ratio.

\item \textbf{Generate $B$ synthetic pattern counts.} For $b = 1,
  \ldots, B$ and each $(y, t)$ stratum, impute the counts
  $\tilde{n}_{j,y,t,1:K}^{(b)}$ of individuals assigned to each
  integration point. The exact imputation model is multinomial with probabilities from \eqref{eq:posterior_pattern_prob}, but
  following the implementation strategy of Section~\ref{sec:sec3}, we
  use the sequential conditional normal approximation described in
  Section~\ref{sec:continuous_relaxation}, with pre-drawn common random
  numbers.

\item \textbf{Compute synthetic summary statistics.} Each subgroup
  analysis partitions the $K$ integration points into distinct groups based
  on a threshold applied to one of the covariates (e.g., weight $> 100$
  kg versus $\leq 100$ kg). For each replicate $b$ and each summary
  statistic $d \in \mathcal{D}_j$, aggregate the synthetic pattern
  counts across the integration points belonging to each subgroup and
  compute $\hat{s}_{j,d}^{(b)}$.

\item \textbf{Form the synthetic likelihood.} Compute $\hat{\mu}_{S,j}$
  and $\hat{\Sigma}_{S,j}$ as the sample mean vector and covariance
  matrix of $\{\hat{s}_{j}^{(b)}\}_{b=1}^B$, and evaluate:
\begin{equation}
f(s_j \mid y_j, \text{trt}_j, \theta) \approx f_{\text{MVN}}(s_j \mid \hat{\mu}_{S,j}, \hat{\Sigma}_{S,j}).
\end{equation}
\end{enumerate}

The complete posterior combines all likelihood components:
\begin{align}
\pi(\theta \mid \text{data})
&\propto \mathcal{L}_{\text{full}}(\theta)
  \times \mathcal{L}_{\text{partial}}(\theta)
  \times \prod_{j \in \mathcal{J}_{\text{BSL}}}
    f_{\text{MVN}}(s_j \mid \hat{\mu}_{S,j}, \hat{\Sigma}_{S,j})
  \times \pi(\theta),
\end{align}
where $\pi(\theta)$ denotes the joint prior on all parameters. The BSL
component therefore acts as an additional source of information.  When the
summary statistics are informative and the synthetic likelihood is
well-calibrated, the BSL-enhanced posterior should therefore concentrate
more tightly around the true parameter values compared to the standard
ML-NMR posterior that uses only the marginalized
likelihood~\eqref{eq:marginal_likelihood}.

Finally, the PSIS correction described in Section~\ref{sec:psis} is
applied post-hoc: in the \texttt{generated quantities} block, we draw
fresh multinomial samples using \texttt{multinomial\_rng}, compute the
discrete synthetic likelihood (using a large value of $B$), and output both log-likelihoods for importance weight calculation.

\section{Application: Plaque Psoriasis Network}
\label{sec:sec5}

We illustrate the BSL-enhanced ML-NMR framework using data from a
network of randomized controlled trials in moderate-to-severe plaque
psoriasis, previously analyzed by \citet{phillippo2020multilevel} and
available in the \texttt{multinma} R package \citep{phillippo2023r}. The binary outcome is achievement of at least 75\% improvement in the
Psoriasis Area and Severity Index (PASI~75) at 12 weeks.  Six treatments are considered: placebo (PBO), ixekizumab Q2W and Q4W (IXE\_Q2W, IXE\_Q4W), etanercept (ETN), and secukinumab 150\,mg and 300\,mg (SEC\_150, SEC\_300).  These treatments are grouped into two classes: IL blockers and TNF$\alpha$ blockers.

\subsection{Network structure}

The network comprises four studies comparing six treatments: placebo
(PBO), ixekizumab Q2W and Q4W (IXE\_Q2W, IXE\_Q4W), etanercept (ETN),
and secukinumab 150\,mg and 300\,mg (SEC\_150, SEC\_300). The
treatments fall into two mechanistic classes: interleukin (IL) blockers
(IXE\_Q2W, IXE\_Q4W, SEC\_150, SEC\_300) and tumor necrosis factor
alpha (TNF$\alpha$) blockers (ETN).  The studies are:

\begin{itemize}
\item \textbf{UNCOVER-1} ($n = 866$): PBO, IXE\_Q2W, IXE\_Q4W.
\item \textbf{UNCOVER-2} ($n = 1,162$): PBO, IXE\_Q2W, IXE\_Q4W, ETN.
\item \textbf{UNCOVER-3} ($n = 1,166$): PBO, IXE\_Q2W, IXE\_Q4W, ETN.
\item \textbf{FIXTURE} ($n = 1,306$): PBO, ETN, SEC\_150, SEC\_300.
\end{itemize}

\noindent Individual patient data (IPD) are available for all three
UNCOVER studies. FIXTURE provides only aggregate (arm-level) data.

\subsection{Covariates and effect modification}

Following \citet{phillippo2020multilevel}, five baseline covariates are included as both prognostic variables and
potential effect modifiers: previous systemic treatment (``prevsys'', binary),
psoriatic arthritis (``psa'', binary), body weight (``weight'', continuous), body
surface area affected (``bsa'', continuous), and duration of psoriasis (``durnpso'', continuous). Also as in \citet{phillippo2020multilevel}, continuous covariates are rescaled: weight is divided by 10, bsa by 100, and durnpso by 10, and all covariates are then centered at their pooled IPD means. This centering ensures that the treatment effect parameters $\gamma_t$ are interpretable as treatment effects at the average covariate values.

We adopt a class-shared effect modification structure (consistent with \citet{phillippo2020multilevel}), where the interaction parameters are common within treatment classes:
$\beta_{2,t} = \beta_{2,c(t)}$ for all treatments $t$ belonging to
class $c(t)$, with $c = 1$ for IL blockers and $c = 2$ for TNF$\alpha$
blockers. This gives two sets of five interaction parameters rather than
five per active treatment. The class-shared assumption is natural here:
within a mechanistic class, treatments share the same target and might
reasonably be expected to exhibit similar patterns of effect
modification across patient subgroups.

For the marginalized likelihood in studies lacking individual covariate
data, we use $K = 64$ study-specific integration points per study,
generated from each study's reported covariate distribution using the
quasi-random integration approach implemented in \texttt{multinma}
\citep{phillippo2023r}. Fixed treatment effects are assumed throughout.

\subsection{Analysis design}
\label{sec:analysis_design}

To evaluate the BSL enhancement, we consider three models that differ in
how UNCOVER-3 is treated:

\begin{enumerate}
\item \textbf{Oracle}: All three UNCOVER studies contribute full IPD
  (i.e., UNCOVER-3 is in $\mathcal{J}_{\text{full}}$). FIXTURE
  contributes aggregate data with marginalized likelihood. This
  represents the best achievable inference.
\item \textbf{ML-NMR}: UNCOVER-1 and UNCOVER-2 contribute full IPD.
  UNCOVER-3 contributes only aggregate counts via marginalized
  likelihood (i.e., UNCOVER-3 is in $\mathcal{J}_{\text{partial}}$).
  FIXTURE contributes aggregate data. Individual covariates from
  UNCOVER-3 are discarded.
\item \textbf{BSL-IS} (BSL with continuous relaxation + PSIS reweighting): Same as ML-NMR, but UNCOVER-3 additionally   contributes subgroup summary statistics via BSL (i.e., UNCOVER-3 is   in $\mathcal{J}_{\text{BSL}}$). The subgroup summaries are computed  from the UNCOVER-3 IPD but are treated as if they were published
  results.
\end{enumerate}

\noindent The comparison between ML-NMR and BSL-IS isolates the information gain from incorporating the subgroup summaries, while the Oracle provides an upper bound. To validate our implementation, we confirmed that both the Oracle and ML-NMR models produce equivalent parameter estimates to those obtained when using the \texttt{multinma} package with matching settings and data.

\subsection{Summary statistics for BSL}

\begin{figure}
    \centering
    \includegraphics[width=0.95\linewidth]{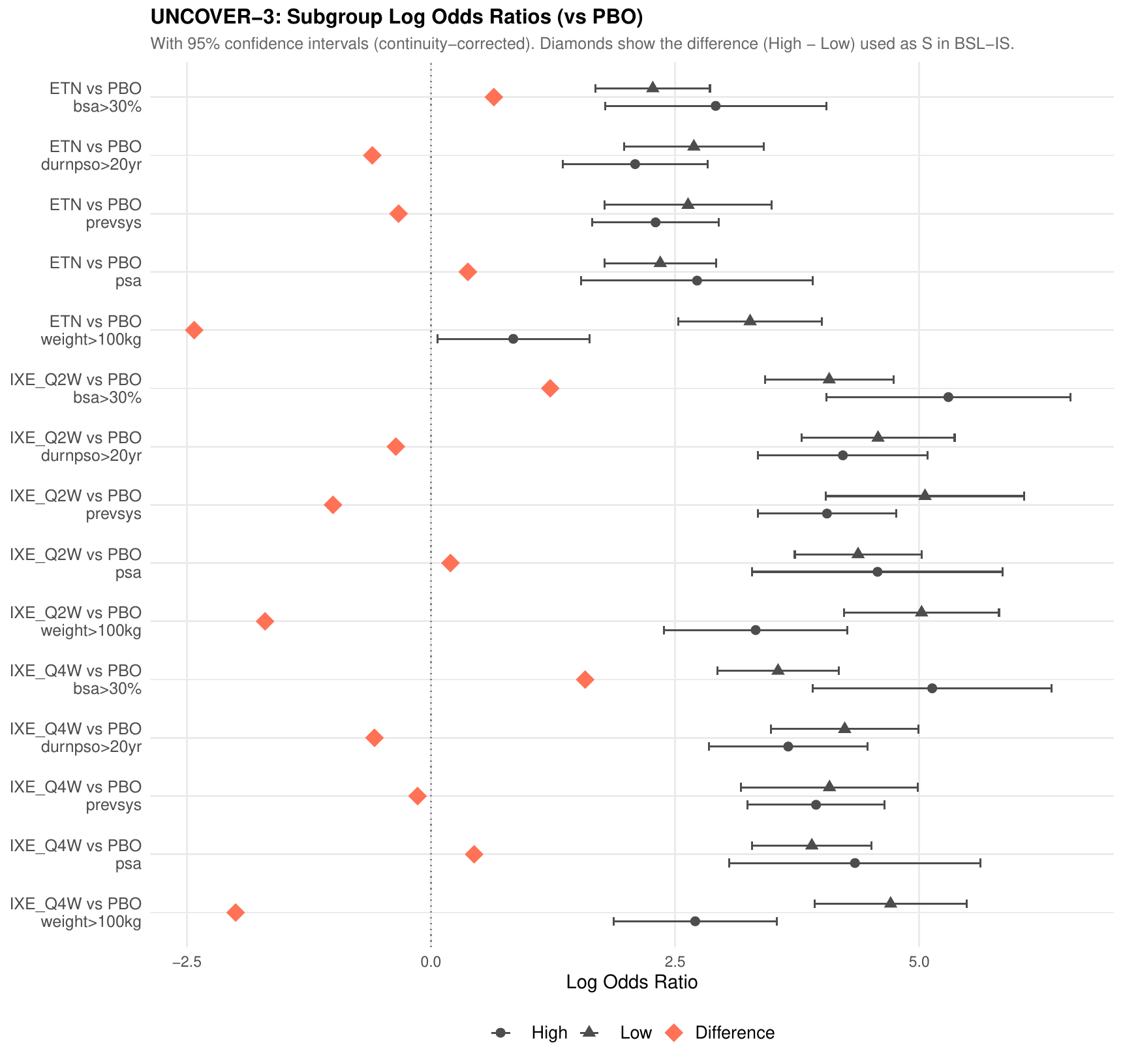}
    \caption{Subgroup-specific log odds ratios (vs.\ PBO) for UNCOVER-3,
    stratified by each covariate split. Points with 95\% confidence
    intervals show the treatment effect within the High (circle) and Low
    (triangle) subgroups. Diamonds show the difference (High~$-$~Low), which
    are the summary statistics used by BSL.}
    \label{fig:summarystats}
\end{figure}

Each of the five covariates defines a binary subgroup split based on a pre-specified threshold (defined on the original, uncentered scale): psa
(0/1), prevsys (0/1), weight ($>$100\,kg vs.\ $\leq$100\,kg), bsa ($>$30\% vs.\ $\leq$30\%), and duration of psoriasis ($>$20 years vs.\ $\leq$20 years). For each split, subgroup-specific log odds ratios (computed with continuity correction) are calculated separately for the ``High'' and ``Low'' subgroups. The BSL summary statistics are then the
\textit{differences} in these log odds ratios across subgroup levels (High~$-$~Low), which directly measure how the treatment effect varies with the splitting covariate. The reference treatment is PBO, so that
comparisons are formed for ETN, IXE\_Q2W, and IXE\_Q4W against PBO, yielding up to $3 \times 5 = 15$ summary statistics. The subgroup log
odds ratios and their differences for UNCOVER-3 are displayed in Figure~\ref{fig:summarystats}.  Subgroup-specific treatment effects with confidence intervals, as shown in Figure~\ref{fig:summarystats}, are routinely reported in clinical
trial publications (often in the supplementary material). Rather than using the subgroup-specific log odds ratios and their confidence intervals directly as summary statistics, which
would yield $\dim(S) = 90$, we use only their differences (High~$-$~Low), reducing the dimensionality to $\dim(S) = 15$ while retaining the most salient information, namely, whether there is evidence of effect modification by each covariate.

\subsection{Results}
\label{sec:results}

To fit the models in the applied example, we used $B = 500$, $K = 64$, $P = 5$, and 4 chains in Stan with $n_{\text{iter}} = 10{,}000$ MCMC iterations (with 5,000 MCMC iterations for burn-in). In the generated quantities block of the BSL-IS model, we used $B_{\text{disc}} = 5{,}001$ (Section~\ref{sec:psis}). We used weakly informative Normal(0,5) priors for all $\gamma$, $\beta_{1}$, and $\beta_{2}$ parameters.  All models showed reasonably good mixing; see trace-plots in Figure~\ref{fig:traceplot} in the Appendix. The computational burden varied substantially across
methods: the Oracle and ML-NMR models required no more than a few minutes, whereas the BSL-IS model required approximately 10 hours.

For the BSL-IS model, the PSIS diagnostic indicated adequate performance with an estimated Pareto $\hat{k}$ of $0.598$. That being said, the effective sample size was $n_{\text{eff}} = 917.1$, suggesting that the analysis may benefit from
either a larger number of MCMC iterations or a larger value of $B$.

\begin{figure}
    \centering
    \includegraphics[width=0.95\linewidth]{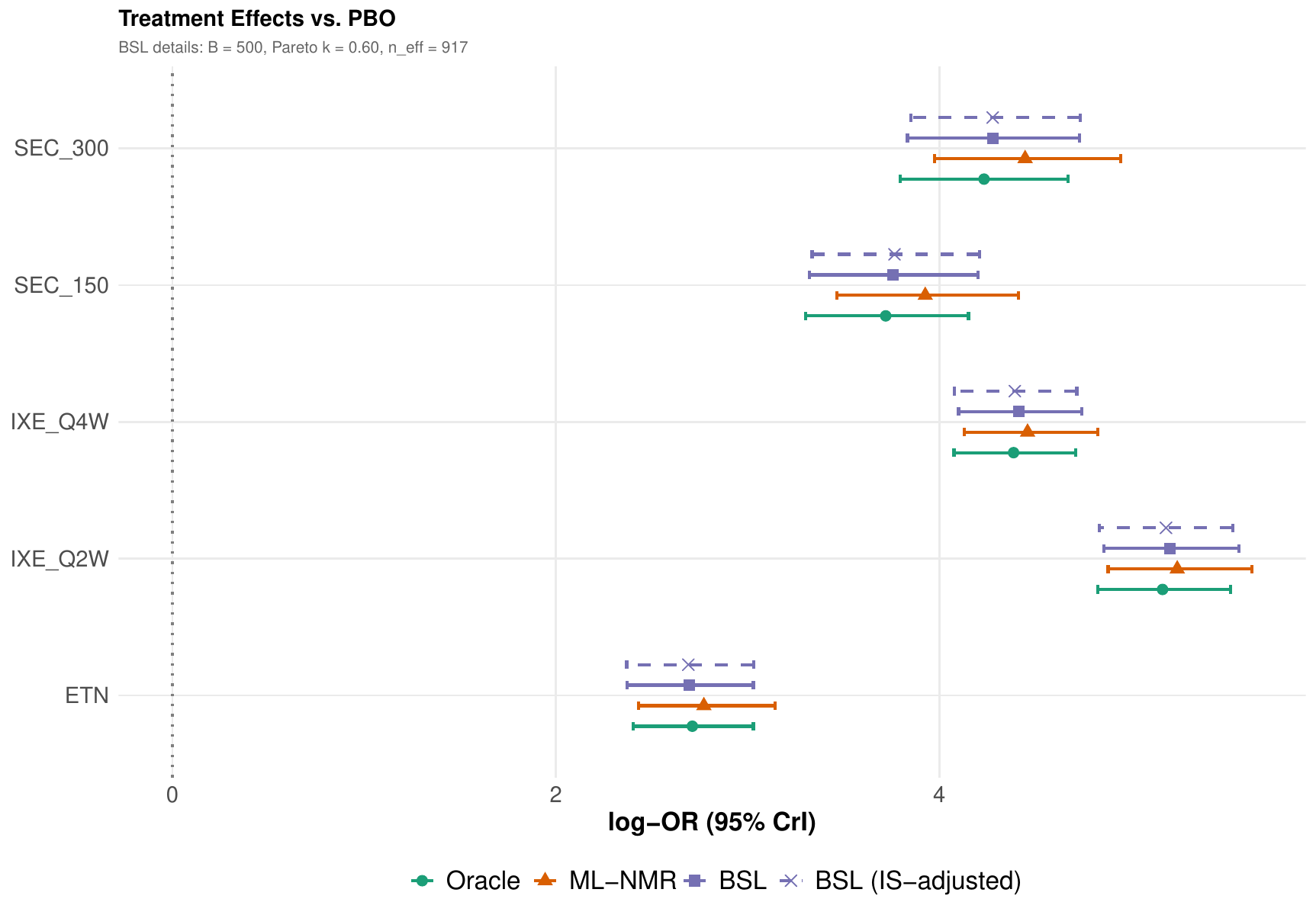}
    \caption{Forest plot displays the posterior mean and 95\% credible intervals for the treatment effect parameters $\gamma_1, \ldots, \gamma_5$ (relative to placebo) under each method.}
    \label{fig:gamma_plot}
\end{figure}

\begin{figure}
    \centering
    \includegraphics[width=0.95\linewidth]{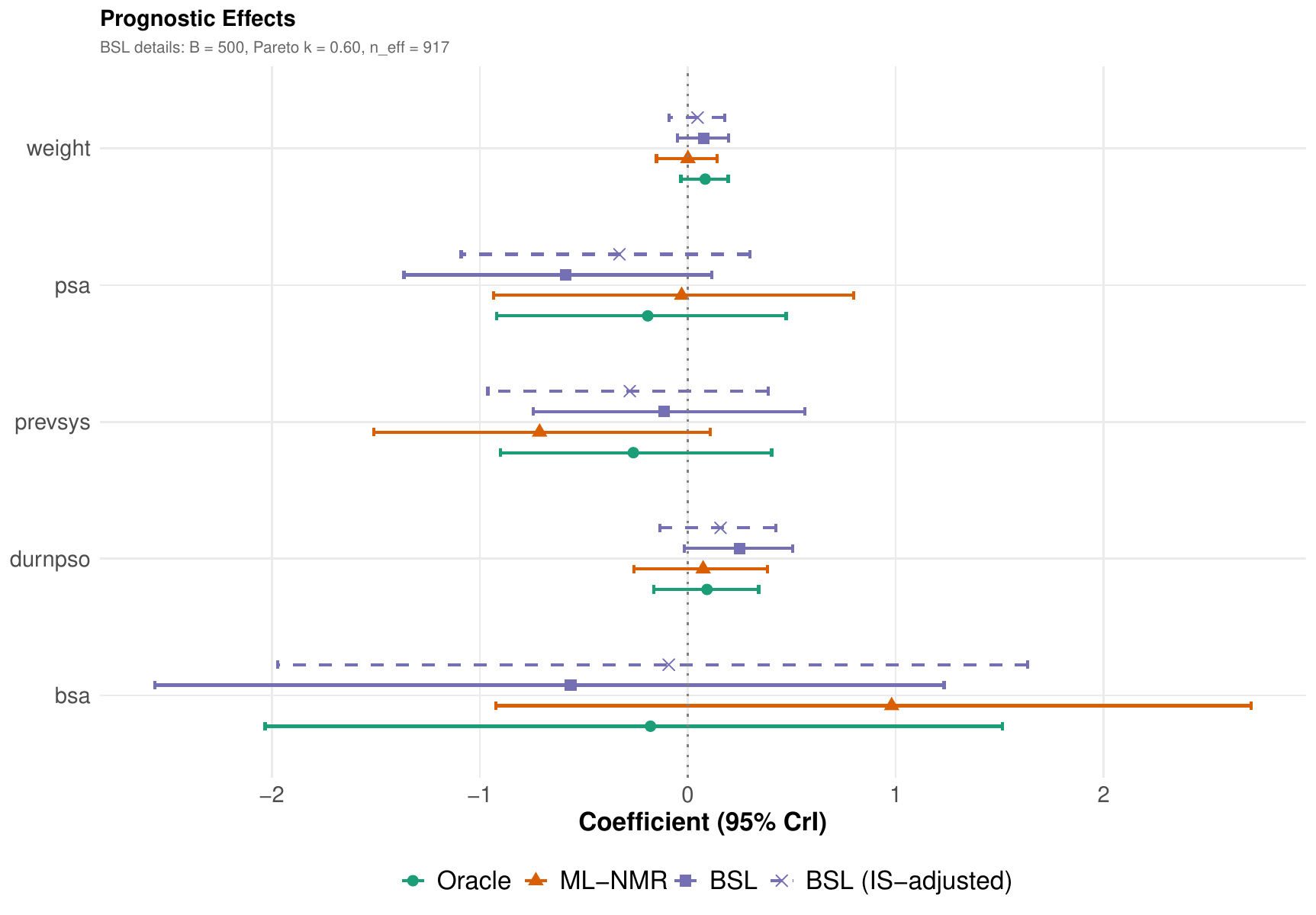}
    \caption{Forest plot displays the posterior mean and 95\% credible intervals for the prognostic effect parameters ($\beta_1$) under each method.}
    \label{fig:beta1_plot}
\end{figure}

\begin{figure}
    \centering
    \includegraphics[width=0.95\linewidth]{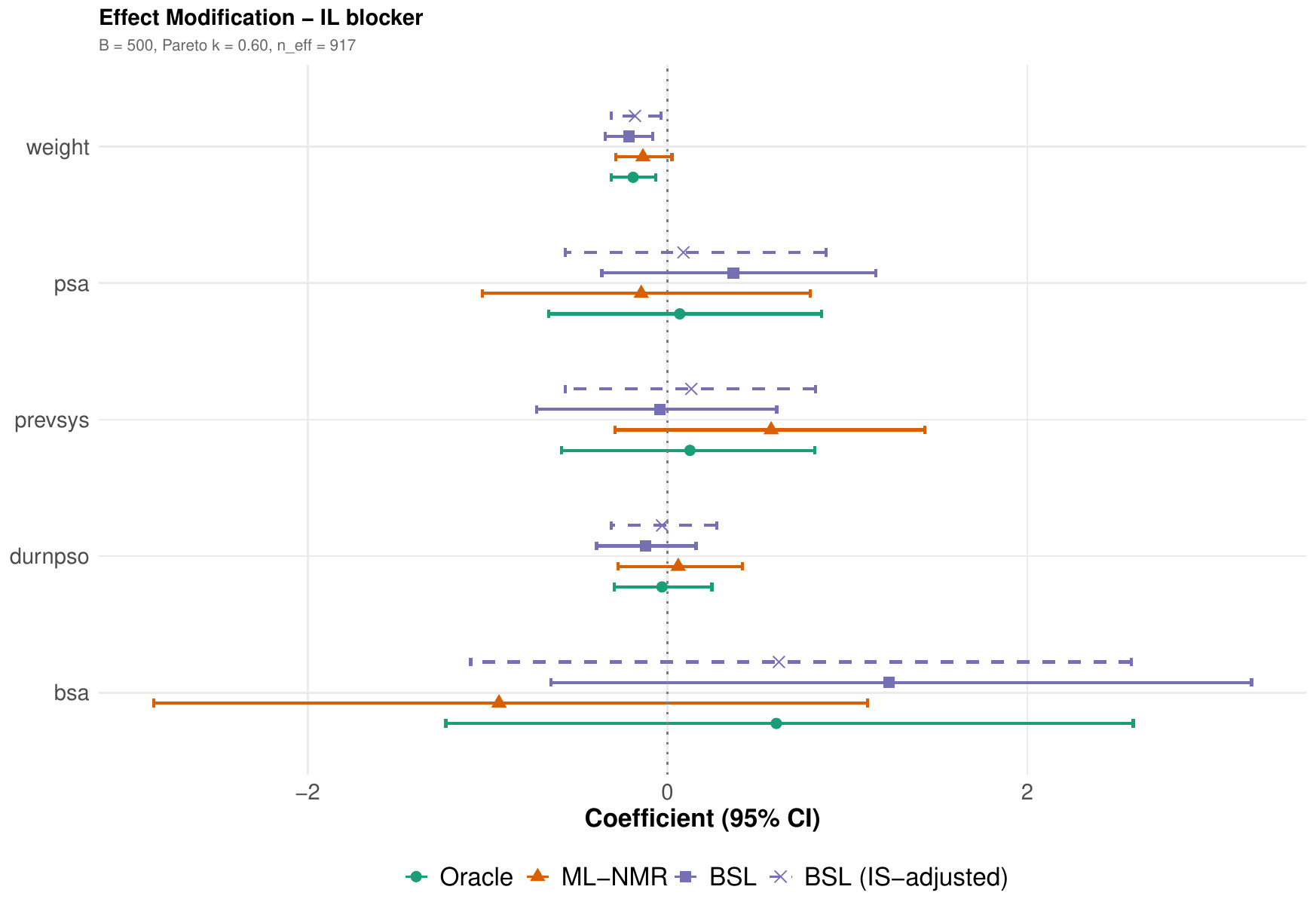}
        \caption{Forest plot displays the posterior mean and 95\% credible intervals for the effect modification parameters ($\beta_2$) for the IL blocker class under each method.}
    \label{fig:beta2B_plot}
\end{figure}

\begin{figure}
    \centering
    \includegraphics[width=0.95\linewidth]{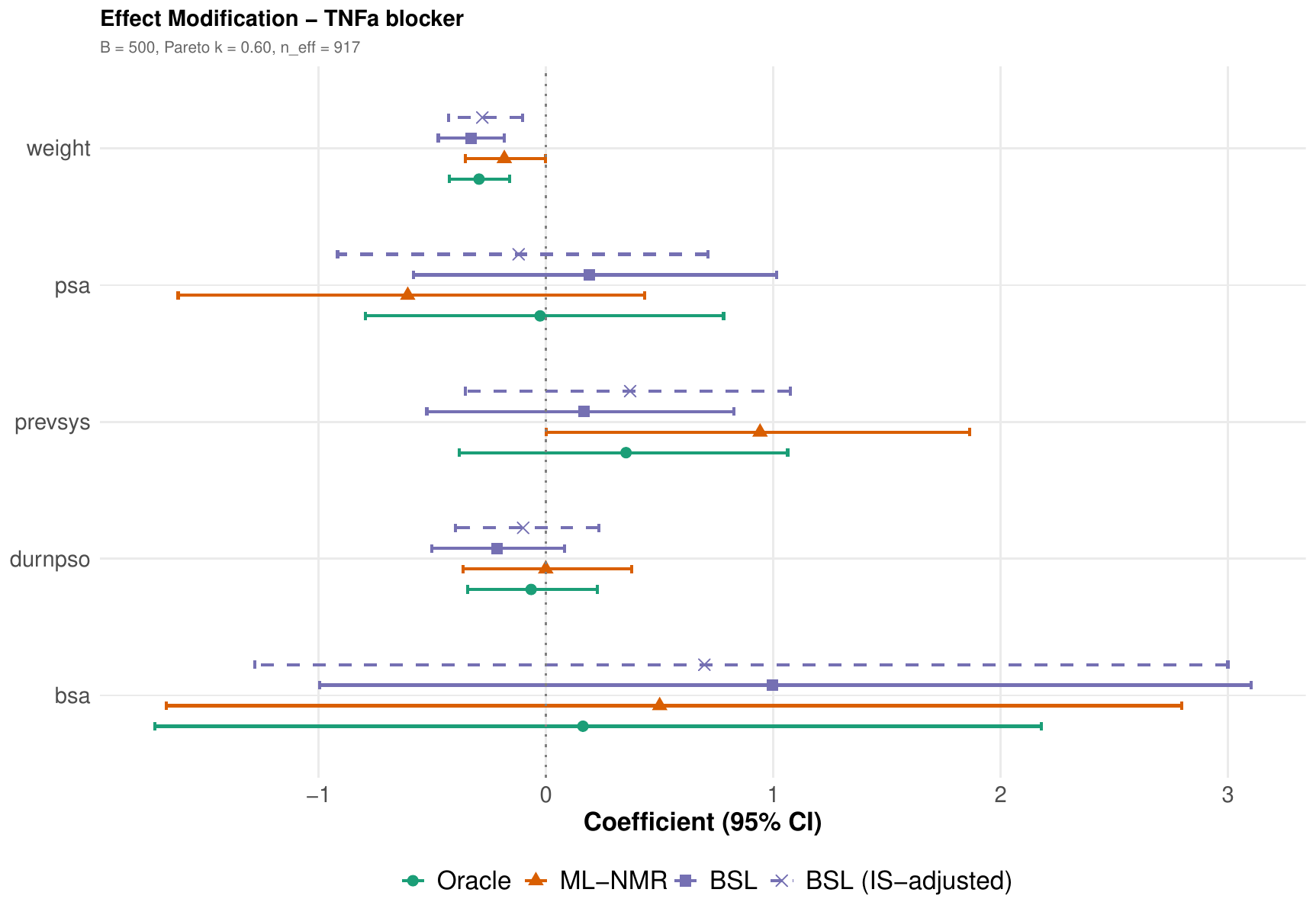}
        \caption{Forest plot displays the posterior mean and 95\% credible intervals for the effect modification parameters ($\beta_2$) for the TNF$\alpha$ blocker class under each method.}
    \label{fig:beta2A_plot}
\end{figure}

Across all five active treatments, the BSL-IS estimates closely tracked the oracle results, in almost all cases much more closely than ML-NMR; see Figures~\ref{fig:gamma_plot}--\ref{fig:beta2A_plot}. For the treatment
effect parameters~$\gamma$, the improvement from BSL-IS over ML-NMR was modest, as ML-NMR already produced estimates close to the oracle in this setting (Figure~\ref{fig:gamma_plot}). The benefit of BSL-IS is more pronounced for the prognostic~($\beta_1$) and effect modification~($\beta_2$) parameters, where ML-NMR showed larger departures from the oracle (Figures~\ref{fig:beta1_plot}--\ref{fig:beta2A_plot}).
This is expected: the subgroup summary statistics directly encode information about how treatment effects vary with covariates, so the synthetic likelihood most directly informs~$\beta_2$, which in turn helps identify~$\beta_1$. 

Among individual parameters, the weight interaction was the most clearly identified effect modifier for both drug classes, with BSL-IS credible intervals excluding zero and closely matching the oracle (IL blocker: $-0.180$, 95\% CrI = ($-0.314, -0.037$); TNF$\alpha$ blocker: $-0.280$, 95\% CrI = ($-0.429, -0.104$)), whereas the corresponding ML-NMR intervals were attenuated toward zero (IL blocker: $-0.136$,  95\% CrI=($-0.288$, $0.025$); TNF$\alpha$ blocker: $-0.183$, 95\% CrI=($-0.356$,$-0.002$)). Conversely, for the previous systemic therapy (``prevsys'') interaction in the TNF$\alpha$ blocker class, ML-NMR produced an estimate of $0.941$ with 95\% CrI=$(0.000, 1.864)$ that borders zero, whereas both the oracle ($0.352$; 95\% CrI =($-0.381$, $1.063$)) and BSL-IS ($0.370$; 95\% CrI =($-0.356, 1.073$)) yielded intervals comfortably spanning zero, suggesting that the apparent effect modification detected by ML-NMR may be an artifact of the information loss from discarding the subgroup data.

Finally, the PSIS adjustment appears to have generally improved BSL estimation. This was most evident with the interaction parameters, where the unadjusted BSL estimates often overshot the oracle, but BSL-IS estimates did not (or at least not to the same extent). This suggests that the continuous relaxation introduces a modest but systematic bias in some parameters, which the importance sampling step is able to largely correct for.

%
%
%

\section{Discussion}
\label{sec:sec6}

We have proposed a BSL approach for incorporating subgroup summary
statistics into ML-NMR, addressing a gap in the population adjustment literature where informative published results are routinely discarded for lack of a tractable likelihood.

We first demonstrated how BSL can be implemented efficiently with HMC in Stan using four complementary strategies: (1)  random numbers to maintain a deterministic target density, (2) sufficient statistic representations to reduce computational cost, (3) continuous relaxation of discrete
distributions to preserve differentiability for HMC, and (4) PSIS to diagnose and correct for the resulting approximation error. These techniques are not specific to the ML-NMR setting and may be useful for other applications where BSL is implemented in gradient-based
probabilistic programming frameworks.  


Using data from a network of plaque psoriasis trials, we then demonstrated that BSL-enhanced ML-NMR can recover much of the information lost when individual-level covariates are unavailable, producing estimates of treatment effects, prognostic coefficients, and
effect modification parameters that closely track those obtained when full individual patient-level data is available. In some cases, the standard ML-NMR analysis yielded qualitatively different conclusions
from the oracle, whereas BSL-IS corrected this discrepancy (e.g., identifying previous systemic therapy as a potential effect modifier for TNF$\alpha$ blockers when the oracle analysis did not). Applying the BSL-IS approach to other datasets with different characteristics (e.g., different sample sizes, numbers of studies, covariate structures, patterns of effect modification, and types of available subgroup statistics) will be important for characterising the settings in which it
offers the greatest gains over standard ML-NMR.  \citet{copas2018role} defined an efficiency measure for quantifying the information gained from incorporating secondary outcomes in multivariate meta-analysis; an analogous measure comparing BSL-IS to standard ML-NMR would help the evaluation of when subgroup summaries are most informative.  In addition to characterising gains in efficiency over standard ML-NMR, of particular interest is whether the additional information from subgroup summaries can lessen the reliance on the shared effect modifier assumption, which is commonly adopted in ML-NMR despite being difficult to justify.

A central challenge in incorporating subgroup summary statistics is that
they are not independent of the individual-level outcome data already
used in the likelihood, so simply appending a marginal likelihood for
these summaries would amount to double-counting. 
For instance, in the applied example it is clear that the subgroup log-odds ratios are deterministic functions of the same outcomes and treatment assignments that enter the marginalised likelihood (i.e.,
$s_{j}$ is not independent of $n_{j,y,t}$). Even if a suitable marginal
likelihood for the subgroup statistics were available in closed form,
deriving the correct \textit{conditional} likelihood (i.e., $p(s \mid y,
\text{Trt}, \theta)$) would be complicated, as it requires
integrating over the joint distribution of the missing covariates
subject to constraints imposed by the observed outcomes (e.g., under the
model, a patient who responded to treatment is more likely to have
covariate values associated with higher treatment efficacy, so the
distribution of covariates conditional on $(y, \text{Trt})$ depends on
the interaction parameters in a complex way). BSL sidesteps this
difficulty: at each MCMC iteration, it imputes the missing covariates by
simulating from the model-implied conditional distribution given the
current parameter values, which are themselves informed by the observed
individual-level data. The synthetic summaries computed from these
imputations therefore automatically reflect the correct conditional
structure, and the multivariate normal approximation to their sampling
distribution serves as a tractable surrogate for the intractable
conditional likelihood.

Reasons for withholding individual-level information about covariates from studies include privacy concerns, intellectual property concerns, and commercial interests of sponsors. Considerable effort has been devoted
to developing methods for federated learning \citep[e.g.,][]{khellaf2024federated} or for generating synthetic individual patient data
that preserve privacy while maintaining analytical utility \citep[e.g.,][]{jiang2025privacy}. Our results suggest a complementary perspective: if sufficiently informative subgroup analyses are
reported, BSL-enhanced ML-NMR may be able to recover much of the information that would have been available from the full individual-level data. Subgroup summary statistics are
inherently privacy-preserving, particularly when minimum cell sizes are enforced. In settings where BSL-IS can closely
approximate the oracle, the practical implication is that publishing
detailed subgroup results may obviate the need for sharing individual-level covariates altogether, at least for the purpose of
population-adjusted indirect treatment comparisons.

For analyses that do require individual-level data, imputation of missing covariate values is an alternative approach. For example,
\citet{zhao2025synthipd} proposed a method which uses
simulated annealing to obtain ``synthetic'' individual patient data based on available aggregate-level data and subgroup summary statistics.  While reconstructing individual-level data is a worthwhile goal, doing so in a way that preserves the inherent correlations and uncertainty is challenging. Exploring how imputation and estimation could be appropriately combined \citep[e.g.,][]{campbell2026fully} is an interesting direction for future work.

Several limitations should be noted. First, the computational cost of BSL is substantial.  In our application, the BSL-IS model required approximately 10 hours to fit compared to minutes for the standard ML-NMR, driven by the need to generate and evaluate $B = 500$ synthetic datasets at each MCMC iteration. The number of synthetic datasets, while large, should perhaps have been larger given the that we obtained a Pareto $\hat{k}$ of $0.598$ (acceptable, but not ideal). As the dimension of~$S$ grows (in our applied example we had $\dim(S) = 15$), so should~$B$, to ensure that $\hat{\boldsymbol{\mu}}_S$ and
$\hat{\boldsymbol{\Sigma}}_S$ are estimated without excessive sampling error.  Accurately estimating $\hat{\boldsymbol{\Sigma}}_S$ as the number of summary statistics grows can be particularly challenging and \citet{priddle2022efficient} have proposed the ``whitening BSL'' approach to address this issue, something we could consider in future implementations of our proposed model along with other possible ways to improve the computational efficiency of BSL \citep{levi2022finding}. While the high computational cost of BSL-enhanced ML-NMR may be acceptable for high-stakes health technology assessments where accurate effect modification estimates are critical, it may limit the feasibility of extensive sensitivity analyses or application to very large networks.

Second, BSL enhancement of ML-NMR is most naturally suited to binary outcomes, where individuals can be classified into a manageable number of categories defined by their outcome and treatment assignment. This classification substantially reduces the complexity of the synthetic data generation step: rather than imputing individual-level covariate values, the BSL procedure operates on category counts via the multinomial procedure described in Section~\ref{sec:sec4}. For continuous or time-to-event outcomes, no such reduction is available and the imputation would need to operate at the individual level, posing considerable computational challenges. Time-to-event outcomes are of particular interest in health technology assessment \citep{cope2023comparison, campbell2025one}, and \citet{phillippo2025multilevel} have recently outlined models to extend ML-NMR to this setting. Enhancing such models to leverage subgroup summary statistics is theoretically possible but may require alternative likelihood-free strategies to remain computationally
practical. Approximate Bayesian Computation (ABC), for instance, uses an acceptance--rejection mechanism that avoids the need to estimate the full covariance matrix of the synthetic summaries, potentially reducing
the per-iteration cost. However, ABC introduces its own difficulties, notably the specification of tolerance parameters and distance metrics, and can perform poorly with high-dimensional summary statistics \citep{frazier2023bayesian, albert2015simulated}.

\textcolor{black}{
Third, we emphasize that regardless of whether indirect treatment comparisons leverage available ancillary subgroup data, they still rely on strong assumptions  regarding the transportability of relative effect measures across a network of evidence. BSL-IS improves the estimation of \textit{observed} effect modifiers, but it does not---and cannot---address unmeasured effect modification. For anchored indirect treatment comparisons (i.e., connected networks of evidence), validity requires the assumption that there are no unmeasured, imbalanced effect modifiers across study populations; and for non-collapsible effect measures, unmeasured prognostic factors may also be problematic \citep{campbell2026noncollapsibility, chandler2026reframing}. In principle, BSL enhancement could also be applied to unanchored indirect treatment comparisons (i.e., disconnected networks of evidence) \citep{chandler2025msr28, beliveau2021theoretical}, provided that one can adequately adjust for all confounders \citep{campbell2025doubly}. However, the stronger identification assumptions required for unanchored comparisons make them inherently more sensitive to model misspecification, and any gains from incorporating subgroup information should not be mistaken for a relaxation of those assumptions.}
 
\textcolor{black}{
On a final and more general note, this work highlights a broad principle: in evidence synthesis problems with partially observed individual-level data, ancillary summary information can be highly informative, and principled methods for incorporating such information deserve greater attention. This principle extends beyond clinical trials: in political polling, for example, Bayesian aggregation models routinely combine topline results from multiple surveys alongside rich aggregate level data (e.g., economic indicators) \citep[e.g.,][]{linzer2013dynamic, heidemanns2020updated} while discarding demographic crosstabs that nearly every poll publishes, despite the rich information these breakdowns contain about how preferences vary across subgroups. As the demand for population-adjusted indirect treatment comparisons continues to grow \citep{serret2023methodological}, approaches that make full use of the available evidence will become increasingly important.}

\bibliographystyle{agsm}
\bibliography{bslmlnmr}

\pagebreak

\section{Appendix}

\begin{figure}
    \centering
    \includegraphics[width=0.90\linewidth]{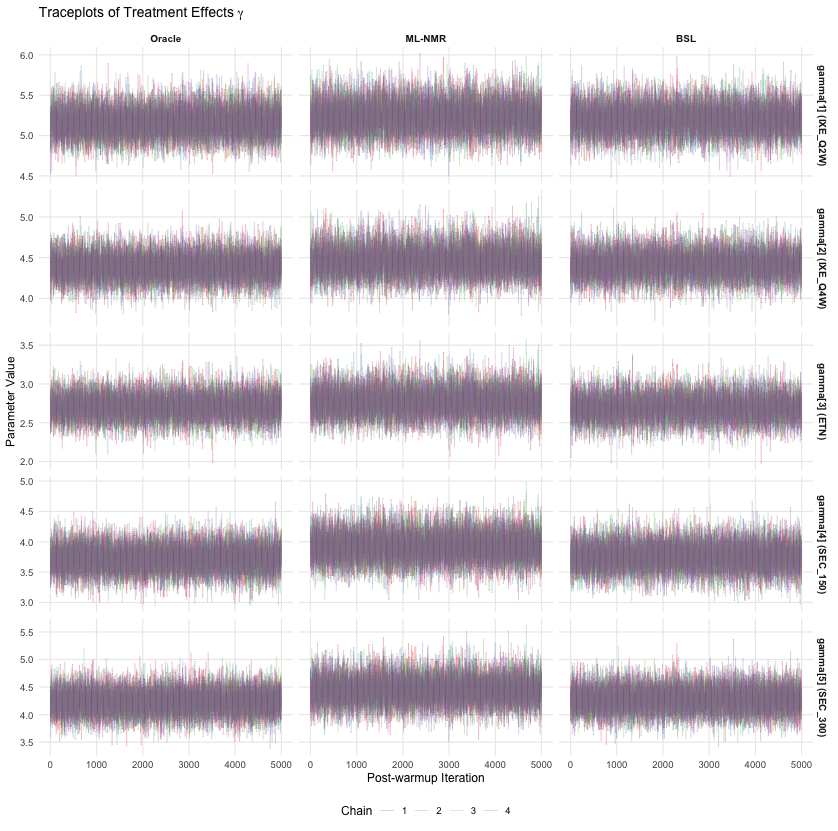}
    \caption{Trace-plots of $\gamma$ for each method, with chains shown in different colors for the Plaque Psoriasis example.}
    \label{fig:traceplot}
\end{figure}

\end{document}